\documentclass[a4]{article}%
\usepackage{amsgen,amstext,amsbsy,amsopn,amsthm, amssymb}
\usepackage{amsmath}
\usepackage{graphicx}
\usepackage{amsfonts}
\usepackage{amssymb}%
\providecommand{\U}[1]{\protect\rule{.1in}{.1in}}
\providecommand{\U}[1]{\protect\rule{.1in}{.1in}}

\begin{document}

\title{Indecomposable semiinfinite string-localized positive energy matter and "darkness"}
\author{Bert Schroer\\CBPF, Rua Dr. Xavier Sigaud 150,\ 22290-180, Rio de Janeiro, Brazil,\\and Institut fuer Theor. Physik der FU Berlin, Germany}
\date{March 2008}
\maketitle

\begin{abstract}
In the absence of interactions indecomposable positive energy quantum matter
comes in form of three families of which the massless so called "infinite
spin" family which appeared first in Wigner's famous 1939 work is (if
mentioned at all) usually dismissed as "unphysical" but without indicating
what principle (if at all) is being violated.

Using novel methods which are particularly suited for problems of
localization, it was shown that these representations cannot be generated by
pointlike localized fields but rather require the introduction of string-like
generators which are localized along semiinfinite spacelike strings. We argue
that such objects can neither be registered in quasilocal Araki-Haag counters
nor pair-produced from standard matter.

Reviewing the mathematical status of Murphy's law in local quantum physics
(everything which is not prohibited to couple does indeed couple) off and on
shell, we are led to the result that perfect darkness is only possible in QFT
with string-localized fields.

\end{abstract}

{Pacs. 95.35+d, 11.10-z, 11.30-Cp }

\section{Localization and darkness\ \ \ }

One of the most enigmatic particle physics enrichments coming from
astrophysical observations is the discovery of "dark matter" in the halos of
galaxies whose contribution to the total matter/energy content of the universe
is a multiple of standard matter. Here by standard matter we mean interacting
theories which are generated by point-like localized fields and which have
positive energy particles of finite spin (helicity) as classified by Wigner
\cite{Wig1}. Even if this new form of matter is not totally inert relative to
standard matter, the consistency with the data demand that at least its
coupling to be weak. The most popular proposals with the least amount of new
parameters came from chargeless components of supersymmetric extensions of the
standard model\footnote{For reasons which we have explained in \cite{infra} we
do not believe that "unparticles" can be a serious contender for DM.}
(neutralinos, WIMPS). Such models lead to measurable production rates of dark
matter through high energy collisions of standard matter, and related
experiments are well on their way.

In this paper the main aim will be to explore the possibility of existence of
perfect dark matter (pDM) in coexistence with standard matter within one model
of local quantum physics. The minimal requirement for perfect darkness is that
the production rate from standard matter vanishes. In a later part of this
article we will also contemplate the possibility that an appropriately defined
algebraic \textit{subspace of counters} (a subspace of the operator algebra of
QFT) for standard matter is inert to pDM. Since the pDM is gravitating, it is
not totally decoupled from standard matter; however we are here primarily
interested in a scenario where even without gravity the word would not tensor
factorize in a standard and a dark part.

Needless to say that pDM will gain astrophysical interest only if those DM
production experiments lead to a null-effect. But since the conceptual setting
is tangent to the core of (especially non Lagrangian) QFT, there is also some
purely theoretical interest even if the here proposed explanation for darkness
will be ruled out by non-vanishing production rate.

It has been known for a long time that properties of confinement and darkness,
which in relativistic QM would be phenomenologically accounted for by
confining potentials or by decoupling production channels, lead to extremely
nontrivial structural problems within local quantum physics.

The main obstacle is the principle of causal locality which has the tendency
of coupling all localized states with each other. This state of affairs is
sometimes referred to as \textit{Murphy's law of local quantum physics}: if
localized states can be coupled (subject to their superselection rules), they
will couple.

As the various sociological versions of Murphy's law, also this one has a
metaphoric connotation since it only expresses an unexpected, ill-understood
tendency. There are various mathematical theorems which may be considered as
its rigorous meaning. One is the Reeh-Schlieder theorem \cite{Haag} which
implies that by operations in a bounded spacetime regions any state in the
universe can be approximated. In fact not only the vacuum, but any finite
energy states is coupled to the rest of the world by any localized algebra, as
small as its localization region may be.

A closely related structural property is the impossibility to decouple
observables localized in a region $\mathcal{O}$ from those localized in the
causal disjoint region $\mathcal{O}^{\prime}$. A special case which shows that
in a local positive energy QFT one cannot decouple a localized projector
$P(\mathcal{O})$ from its spacelike separated translate is Malament's theorem
\cite{Mal}. Although not directly related to the issue of this paper (but
related to Malament's theorem) we mention in passing that the perhaps most
spectacular disparities between relativistic\footnote{By relativistic QM we
mean the Poincare representation theoretical setting of \textit{direct
particle interactions} which leads to a Poincare invariant clustering S-matrix
\cite{CP} but cannot implement the local covariance principle which is
characteristic for QFT.} QM and local quantum physics (LQP) shows up in the
two different notions of entanglement based on the two different localization
concepts (the Born-Newton-Wigner localization in QM and the modular
localization in LQP \cite{MSY}). The insufficient distinction between the
information theoretic quantum mechanical entanglement from the thermal
manifestation \cite{S3} (localization entropy \cite{S2}) has been the cause of
many confusions, including those behind the information loss in black hole physics.

In some way the split property \cite{Haag}, by which one can enforce a tensor
factorization after \textit{separating }$\mathcal{O}$\textit{ from
}$\mathcal{O}$\textit{' by creating a attenuation distance }$e$\textit{ for
the vacuum fluctuations }$\mathcal{O}_{\varepsilon}^{\prime}\subset
\mathcal{O}^{\prime}$ \cite{S2}, restores some of the properties of QM. but
the result is still not the ideal world of QM since all states which are
relevant in particle physics (vacuum, finite energy states) become
\textit{thermally entangled} states with respect to a new Hamiltonian
associated with the localization region. On the other hand the factorizing
states have unbounded energy and particle number and hence lack a clear
physical interpretation in terms of particles.

The root of all these unusual manifestations is the radically different
structure of the local algebras: whereas in (relativistic) QM the B-N-W
localized algebras are algebras of bounded operators B(H) in a factor Hilbert
space, the modular localized algebras are copies of the monad algebra (the
unique hyperfinite factor of type III$_{1}$) \cite{S3}. This algebraic
distinction accounts for all the structural differences.

Hence in contrast to Murphy's law in everyday life, its LQP atavar is not only
the cause of complicating the quantum mechanical life by coupling infinitely
many channels, but it also creates that conceptual tightness which accounts
for the fact that we consider QFT in comparison with relativistic QM as the
more fundamental setting. The before mentioned Leibnitz picture of reality in
LQP which states that the complex structure of QFT can be encoded into the
relative positioning of a finite number (2 in chiral models, 3 in d=1+2 , 6 in
d=1+3,...) of structureless "monads" without any individuality (isomorphic to
the unique hyperfinite III$_{1}$ factor) \ is inexorably linked to the
apparent messy aspects which LQP presents from a quantum mechanical setting.

Having exposed some of the conceptual problems which one has to confront
before envisaging darkness in the above perfect sense, we now ask the more
concrete and observational relevant question \textit{whether the S-matrix can
escape the on shell adaptation of Murphy's law}. Using on-shell analytic
properties which extend those which one can derive in the presence of
interpolating local fields (in particular the crossing relation), one can show
that a pure elastic scattering without the participation of creation channels
is impossible. This nonperturbative assertion known as Aks' theorem
\cite{Aks}, which with the passing of time has acquired the Rumsfeldian status
of an "unknown known", is limited to 4 and higher dimensions. In lower
dimension, in particular in d=1+1 there is a subterfuge which is related to
the Coleman-Mandula issue arising from infinitely many conservation laws. The
localization aspect behind this mechanism is the existence of rather
well-behaved ("temperate") vacuum-\textbf{p}olarization-\textbf{f}ree
one-particle \textbf{g}enerators (PFG's) of the wedge algebra in
d=1+1\footnote{Any smaller localization region does not permit the existence
PFG's in the presence of (any) interaction.} \cite{BBS}.

The Aks' result would in particular exclude a vanishing production rate of DM
from standard matter, at least in case that the DM is (as the SM) described by
pointlike localized fields. But the principles of QFT permit field algebras
which cannot be generated by fields which are better localized than
semiinfinite string-like. Here the observables are always assumed to be
point-like generated i.e. the candidates for string-like localized generating
fields\footnote{It is regrettable that the good autonomous use of this
terminology always requires one to add that its use in string theory is
totally metaphoric; e. g. the N-G quantum object is a generalized free field
\cite{crisis}.} carry nontrivial "Casimir charges" which distinguish the
string-generated sector from the vacuum sector.

String-like localization is the most general localization which follows from a
mass gap spectrum and the derivation of this fact by Buchholz and Fredenhagen
\cite{BF} belongs to one of the most profound conceptional enrichments of QFT.
The stringlike localization destroys the possibility of describing the same
out/in configuration by the $t\rightarrow\pm$ limits of the same wave-packet
smeared Heisenberg field; rather the string direction in both cases must be
chosen different in order to obtain the same out/in configuation \cite{BF}.
With the breakdown of crossing the main obstacle leading to the Aks theorem
has been taken out and with it the on-shell reign of Murphy's law has been
weakened. In fact in the conclusions of \cite{BF} one encounters the remark
that a hypothetical string quark matter, once bound into hadrons or lost into
the surrounding space, cannot be produced because the string-localization may
impede the production process. Nowadays we would think that this idea works
better with DM.

Semiinfinite string-like localization may be incompatible with Lagrangian
quantization, but it is the kind of localization which can be rigorously
derived from the mass gap assumption in conjunction with local observables. In
such a setting the full field algebra is generated by string-localized (basic)
fields which contains a subalgebra which is generated by pointlike composites.
In order to create all the particles from the vacuum one needs the string
fields which in some way play a similar role as the interpolating fields in
the LSZ scattering theory\footnote{Strictly speaking the stringlike
localization destroys the possibility of describing the same out/in
configuration by the $t\rightarrow\pm$ limits of the same wave-packet smeared
Heisenberg field; rather the string direction in both cases must be chosen
different in order to obtain the same out/in configuation \cite{BF}. This is
at the basis of the previously mentioned breakdown of crossing.}. localization
which follows from massive theories with a mass gap. There exists at present
no model which requires string-localized fields, but given the lack of
nonperturbative knowledge even with respect to point-like localized Lagrangian
models, their existence can hardly be doubted. The observable imprints such
stringlike fields leave on the level of particles and their interactions are
rather subtle; the particle spectrum with gaps just looks the same as for
point-like fields and the observable content of a breakdown of the analytic
aspects of on-shell crossing\footnote{The conceptual change is however quite
drastic: the uniqueness argument in \cite{unique} for the inverse scattering
which uses crossing will break down. The one S-matrix one QFT inverse
scattering statement has bee first observed in factorizing models. The
validity of crossing extends the uniqueness (but not the existence) to higher
dimensions.} is hard to assess. However the main ingredient into Aks' proof is
the validity of crossing and the absence of certain creation channels would
perhaps constitute the most dramatic manifestation of the presence od
string-like generators.

An interesting development which presently drives a good part of structural
research about localization properties and in particular the use of lesser
localized objects as wedge-localized generators is the interesting fact that
among those objects one can find generators with significantly simpler
properties \cite{Ann} than among interacting point-like fields. The latter
create, as already alluded to before, in the presence of (any) interactions
upon application onto the vacuum very complicated infinite vacuum polarization
clouds interaction induced In this way the first hard mathematical existence
theorems about an interesting strictly renormalizable (beyond the
superrenormalizable models of the 60ies) models have been proven and the
(generally hard) problem of asymptotic completeness was established
\cite{Lech}. The mathematical concepts applied to d=1+1 has not yet led to an
existence proof of higher dimensional QFT but an interesting link with
attempts to construct nonlocal QFT of the Moyal deformation kind has arisen
\cite{GL}. This is another attempt to cut a breach into the territory governed
by the on-shell Murphy's law.

In this note we want to direct the reader's attention to an even more radical
type of perfectly invisible matter i.e. quantum matter which remains either
totally or partially inert relative to standard matter. As all positive energy
matter it has a gravitational manifestation. But different from standard
matter, and also from the previously presented massive strings, its
semiinfinite string nature of its field generators leads to indecomposable
semiinfinite string-localized states. Clearly this is only possible in the
presence of zero mass; indeed such irreducible zero mass representation do
occur in the Wigner's classification of indecomposible positive energy
solutions and in the next section we will remind the reader of their
properties. This extreme form of hypothetical matter cannot even trigger
Araki-Haag counters \cite{Haag}. A-H counters are counters of localizable
quantum matter represented by operators which do not respond to the vacuum but
register to excitations above the vacuum; as a result of the Reeh-Schlieder
theorem they cannot be strictly local but they are contained in the subalgebra
of \textit{quasilocal} operators i.e. operators which admit a \textit{rapid}
approximation by local operators\footnote{The analysis with A-H counters which
leads to a generalization of the Wigner particle theory ("particle weights"
\cite{Bu-Po}) is primarily a theoretical concept since it ignores the fact
that the final registration needs the intervention of light emitted by charged
particles.}.

\section{String-localized states from representation theory}

The role of semiinfinite string localization as the best generators for those
representations of the Poincare group which are inconsistent with point-like
generators was not seen for a very long time. This had two quite interesting
historical reasons. On the one hand Wigner used a localization concept (the
Born-Newton-Wigner localization) which is obtained from quantum mechanics by
adjusting the localization operator $\vec{x}_{op}$ to the L-invariant inner
product, thus maintaining the probability aspect and its relation to
projectors in the relativistic context but lacking local covariance; in a
quantum theory without a maximal velocity the independence from a reference
frame is only restored asymptotically between asymptotic timelike separated
events\footnote{This fact is of crucial importance for the S-matrix in
relativistic QM and for the probability interpretation of the S-matrix in
terms of cross sections.} whereas the principle of local covariance in QT with
a maximal velocity requires the \textit{modular localization} (which Wigner
missed and which will be explained below). So on the bright side the two
important but very different localization concepts coalesce asymptotically, a
fact which is crucial for the Poincare invariance and the probability
interpretation of the S-matrix. Although particles in interacting QFT do not
exist, their asymptotic appearance accounts for the observable richness of
particle physics.

Besides Wigner's missing the appropriate covariant localization concept for
finite propagations, there is a second reason why string localization posed a
difficult conceptual hurdle. QFT was discovered by Jordan in the form of
quantization i.e. a parallelism of the classical field formalism. But it turns
out that this parallelism is limited to pointlike fields \textit{since
indecomposable string-like quantum states have no classical analog and
classical relativistic strings have an associated QFT which is not
string-localized in any intrinsic quantum sense}. This can be explicitly
illustrated by realizing that on the one hand the class of infinite spin
representations do not arise from a classical Lagrangian and on the other hand
To illustrate this point in a completely explicit manner: the Wigner infinite
spin representation does not arise from quantizing a classical Lagrangian and
the quantum object associated with the classical Nambu-Goto string is not a
quantum string-localized field but a point-like so-called generalized free
field \cite{crisis} i.e. the interpretation of this localization point as the
center of mass point of a quantum string is a metaphoric invention which
ignores the intrinsicness of the quantum localization concept\footnote{Not all
treatments of the quantum N-G model have followed that metaphor suggested by
Euclidean functional representation. In \cite{Pohl} the quantization was done
by using the complete integrability of the model and it was later shown that
the resulting theory is not equivalent to the canonically quantized one
\cite{Bahns}.
\par
{}}.

Wigner found that there are precisely three families of indecomposable
positive energy representations of the Poincare group, two rather large
families containing continuously many inequivalent representation and one
representation family which is of countable cardinality. They are
distinguished by the nature of the little group and its representation theory.

Besides the best studied massive family, for which the little group is the
invariance group of a timelike vector (and hence isomorphic to $SO(3)$), there
exist two massless families whose little group of a lightlike vector is
isomorphic to a noncompact euclidean subgroup group of the Lorentz group
$E(2)\subset L(3,1)$. Since the representation of the P-group is induced from
$E(2)$, this property is passed on to the $P$-representation.

What distinguishes the two massless families is the \textit{nature of the
}$E(2)$\textit{ representations}; whereas the finite helicity family which
contains the known zero mass particles is a degenerate representation in which
the euclidean translation is represented trivially (which compactifies the
representation despite the noncompactness of the group), the third family
results from a faithful $E(2)$ representation which preserves the group
theoretic noncompactness and comes with unusual, conceptually challenging
properties. The little Hilbert space is now an infinite dimensional space of
Fourier components which describe an $E(2)$-irreducible infinite intrinsic
abelian angular momentum tower; this is why we prefer the terminology
"infinite spin" over Wigner's "continuous spin" (which refers to the
continuous values of the Casimir invariant). The appearance of this infinite
tower prevents the extension of the P-group to the conformal group despite the
vanishing of the mass.

These positive energy representation of the third kind had a long and
complicated history; most particle theorists, including Wigner and later
Weinberg felt that there is something unphysical about these representation
i.e. a conceptual reason why nature apparently only uses the two other classes
of positive energy representations but not this one. The first theorem showing
that these representations do not allow pointlike field generators appeared in
1970. The problem of whether these representations have some residual
localization lay dorment up to recently \cite{BGL} when it was shown that
spacelike cone localization of states is a consequence of the positive energy
condition\footnote{The reader will notice a similarity to the B-F theory of
massive strings. But the latter are operator strings whose action on the
vacuum creates massive states which have a pointlike decomposition theory
under the Poincare group whereas the massless third kind Wigner states are
indecomposable string-like i.e. the state decomposition theory goes beyond the
generating theory of operator algebras.}. It turned out \cite{MSY1} that the
local covariant generators of such representations are string-localized fields
$\Phi(x,e)$ localized on semiinfinite spacelike lines $x+\mathbb{R}%
_{+}e,~e^{2}=-1$

The mathematical framework of the relevant quantum localization concept is
fairly new and goes under the name of \textit{modular localization}%
$~$\cite{Ann}\cite{BGL}.

Since in the deafening noise of present particle physics fashions progress on
old conceptual problems are hardly noticed, we sketch the main idea of modular
localization without proof in the simplest spinless case (where also
traditional methods would be sufficient) and only quote the results for the
case at hand. For more details we refer to the mentioned literature.

Intuitively modular localization is the quantum counterpart of causal
localization in classical field theories with a maximal propagation speed. It
is inherent in relativistic QFT and its structural quantum properties become
more exposed after one liberates it from the use of particular field
coordinatizations. In other words modular localization is the standard
localization implemented by (necessarily singular) field coordinates in
relativistic QFT after one separates the unique localization concept from the
highly non-unique coordinatization-dependent aspects of field generators.
Hence modular localization is a property of the system of local algebras which
does not depend on which field among the infinite number of possible field
coordinatizations one selects.

Here we are interested to localize states in a Wigner representaion. Starting
from a space of wave functions of a scalar particle one first defines two
commuting operators in $H_{Wig}$ which are associated to the $t-x$ wedge
$W_{0}=\left\{  x~|~x_{1}>\left\vert x_{0}\right\vert \right\}  ,$ namely the
unitary representers $\mathfrak{u}$ of the wedge-preserving Lorentz boost
$\Lambda_{W_{0}}(\chi)$ which commutes with the antiunitary representer
$\mathfrak{u}(j_{W_{0}})$ of the wedge-reversing reflection $j_{W_{0}}$ across
the edge of the wedge (third line).%

\begin{align}
&  H_{Wig}=\left\{  \psi(p)\ |~\int\left\vert \psi(p)\right\vert ^{2}%
d\mu(p)<\infty\right\} \\
&  \left(  \mathfrak{u}(\Lambda,a)\psi\right)  (p)=e^{ipa}\psi(\Lambda
^{-1}p),\nonumber\\
&  \left(  \mathfrak{u}(j_{W_{0}})\psi\right)  (p)=\overline{\psi(-j_{W_{0}%
}p)}\nonumber
\end{align}
One then forms the \footnote{The unboundedness of the $\mathfrak{s}$
involution is of crucial importance for the encoding of geometry into domain
properties of unbounded operators.} \textquotedblleft analytic
continuation\textquotedblright\ in the rapidity $\mathfrak{u}(\chi
\rightarrow-i\pi)$ which leads to unbounded positive operators. Using the
customary notation in modular theory, one passes to the following unbounded
closed antilinear involutive operators in $H_{Wig1}$
\begin{align}
&  \mathfrak{s(}W_{0}\mathfrak{):}=\mathfrak{\ \mathfrak{j}}_{W_{0}}%
\delta_{W_{0}}^{\frac{1}{2}},~~\delta_{W_{0}}^{it}:=\mathfrak{u}_{W_{0}}%
(\chi=-2\pi t)~\label{pol}\\
\mathfrak{\ }  &  \left(  \mathfrak{s(}W_{0}\mathfrak{)}\psi\right)
(p)=\psi(-p)^{\ast},~dom~\mathfrak{s(}W_{0})\ \mathfrak{=~}dom\ \delta_{W_{0}%
}^{\frac{1}{2}}\nonumber
\end{align}
where the analytic properties of the domain of this unbounded modular
involution $\mathfrak{s(}W_{0}\mathfrak{)}$ with $\mathfrak{s}^{2}%
\mathfrak{(}W_{0}\mathfrak{)\subset~}\mathbf{1}$ consists precisely of that
subspace of Wigner wave functions which permit that analytic continuation on
the complex mass shell which is necessary in order to get from the forward to
the backward hyperboloid ($\chi\rightarrow\chi-\pi i$). The main assertion of
modular localization is that the $\pm1$ eigenspaces (real since $\mathfrak{s(}%
W_{0}\mathfrak{)}$ is antiunitary) are the real closed component of the dense
$dom$ $\mathfrak{s(}W_{0})$%
\begin{align}
\mathfrak{K}(W_{0})  &  =\left\{  \psi|~\mathfrak{s(}W_{0}\mathfrak{)}%
\psi=\psi\right\}  ,~\mathfrak{s(}W_{0}\mathfrak{)}i\psi=-i\psi\\
&  dom~\mathfrak{s(}W_{0}\mathfrak{)}\mathfrak{=K}(W_{0})+i\mathfrak{K}%
(W_{0})~\nonumber\\
&  \mathfrak{s(}W_{0}\mathfrak{)(\psi+}i\varphi)=\mathfrak{\psi-}%
i\varphi\nonumber
\end{align}
The dense subspace $dom~\mathfrak{s(}W_{0}\mathfrak{)~}$(i.e. $\overline
{dom~\mathfrak{s(}W_{0}\mathfrak{)}}=H_{Wig}$)$\mathfrak{~}$is precisely the
one-particle component of the$~W_{0}$ localization space associated with a
scalar free field $A(x)$, or in terms of the real subspace\footnote{The
closedness of $K$ does not imply that of $K+iK.$}%
\begin{equation}
\mathfrak{K}(W_{0})=clos\left\{  (A(f)+A(f)^{\ast})\Omega~|~\sup pf\subset
W_{0}\right\}
\end{equation}
but the modular construction of localized subspaces avoids the use of singular
field coordinatizations and the ensuing smearing with classically localized
test functions and relies instead on the more intrinsic description in terms
of domains of distinguished unbounded operators in the \footnote{It was
precisely this uniqueness which was Wigner's main motivation for bypassing the
confusing plurality of the quantization setting (many different equations of
motion with the same physical content) in favor of an intrinsic description.
However the adaptation of the Born particle localization (the Newton-Wigner
localization) was taking him away from covariant causal locality.} Wigner
space associated with the representation $(m,s=0)$. The second line is the
defining relation of what is called a \textit{standard real subspace }of a
Hilbert space\textit{. }The \textit{standardness property} is equivalent to
the existence of an abstract (nongeometric) modular involution.

Applying Poincar\'{e} transformations one generates from $\mathfrak{s(}%
W_{0}\mathfrak{)}$ and$~\mathfrak{K}(W_{0})$ to the $W$-indexed families
$\left\{  \mathfrak{s(}W\mathfrak{)}\right\}  _{W\in\mathcal{W}},~\left\{
\mathfrak{K}(W)\right\}  _{W\in\mathcal{W}}.$ The localization spaces for
smaller causally complete spacetime regions $\mathcal{O}$ (which could be
trivial) are obtained by intersections $\mathfrak{K}(\mathcal{O}%
)=\cap_{W\supset\mathcal{O}}\mathfrak{K}(W).$ A remarkable property of all
these spaces resulting from Wigner%
\'{}%
s positive energy representation setting is the validity of \textit{Haag
duality}
\begin{equation}
\mathfrak{K}(\mathcal{O}^{\prime})=\mathfrak{K}(\mathcal{O})^{\prime}
\label{Haag}%
\end{equation}
where the dash on the region denotes the causal complement and that on the
K-space stands for its symplectic complement within $H_{Wig}$ i. e.
$Im(K,\varphi)=0$ for all $\varphi\in\mathcal{K}(\mathcal{O})^{\prime
}=j_{\mathcal{O}}\mathfrak{K}(\mathcal{O})$

The final step is the functorial ascend to the net of spacetime localized
operator algebras in the Wigner-Fock space $H_{W-F}=H_{F}(H_{Wig})$ (with
creation/annihilation operators $a^{\ast}(p),a(p)$)%
\begin{align}
&  Weyl(\psi)=\exp i(a(\psi)+a^{\ast}(\psi)),~\psi\in\mathfrak{K}%
(\mathcal{O})\\
\mathcal{A(O)}  &  :=alg\left\{  Weyl(\psi)~|~\psi\in\mathfrak{K}%
(\mathcal{O})\right\}  ,~\mathcal{A}:=\cup_{\mathcal{O}}\mathcal{A(O}%
)\nonumber\\
K(\mathcal{O})  &  =\overline{\left\{  \left(  A+A^{\ast}\right)
\Omega\ |\ A\in\mathcal{A}(\mathcal{O})\right\}  },~\mathfrak{K}%
(\mathcal{O})=P_{1}K(\mathcal{O})\nonumber
\end{align}
where $alg$ denotes the operator (von Neumann) algebra generated by the
unitary Weyl operators in the Wigner-Weyl space and $P_{1}$ is the projection
of the Wigner-Fock space onto the Wigner one particle space. Note that there
are no spacetime dependent field coordinates, the construction is as intrinsic
and unique as the Wigner representation theory.

This modular construction exists for all three Wigner representation families.
The $\mathfrak{K}(\mathcal{O})+i\mathfrak{K}(\mathcal{O})$ spaces for
$\mathcal{O=D=}$ double cone (the prototype of a simply connected causally
complete compact region) for the first 2 families are dense in $H_{Wig}$
whereas the third kind of Wigner matter yields a vanishing $\mathfrak{K}%
(\mathcal{D}).$for double cones. In that case the nontrivial space with the
tightest localization $\mathfrak{K}(\mathcal{C})$ is associated with an
(arbitrarily thin) noncompact spacelike cone $\mathcal{C}$ $=x+\mathbb{R}%
_{+}\mathcal{D}$ with apex $x$ and an opening angle which is determined by
$\mathcal{D}.$ All relations about $\mathfrak{K}$ pass to the K's in
Wigner-Fock space.

There is no problem in adapting the modular setting to the presence of
interactions; however different from the free situation there are no
one-particle creators in compactly localized algebras (for details see
\cite{BBS}).

The steps explained above in the spinless context can be carried out for the
first two families with the help of intertwiners. These can also be
constructed without modular theory by standard group theoretical techniques as
explained in Weinberg's first volume of \cite{Wei}. They intertwine the unique
Wigner representation to the denumerable infinite set of $\left(  2A+1\right)
(2\dot{B}+1)$ component \textit{spinorial fields} indexed by $r=(A,\dot{B})$
\begin{align}
&  \Phi_{r}(x)=\sum_{k=-s}^{s}\int d\mu(p)\{e^{ipx}u_{k,r}(p)a^{\ast
}(p,r)+\nonumber\\
&  +e^{-ipx}u_{c}(p)_{k,r}b(p,r)\},~\left\vert A-\dot{B}\right\vert \leq s\leq
A+\dot{B}: \label{inequ}%
\end{align}
But only in the massive case the full spinorial range ($A,\dot{B}$) relative
to the given Wigner spin s is realized; the massless case, as a result of its
different little group requires in addition $A-\dot{B}=0.$ This generates big
gaps in the full spinorial spectrum \footnote{There are more subtle
differences to the massive case whose exploration requires more future
research: whereas massive representations are Haag dual (\ref{Haag}) for
spacetime regions of arbitrary higher connectivity, this is not so for the
massless representations starting from helicity one HD does not extend beyond
simply connected regions.}. In particular there is no covariant
vectorpotential for $s=1$ and no metric potential $g_{\mu\nu}$ for $s=2;$ in
both cases one has to go to higher degree tensor fields than in the massive
case (the field strengths $F_{\mu\nu}$ and the linearized Riemann tensor
$R_{\mu\nu\kappa\lambda}$) in order to remain within quantum physics. On the
other hand a covariant semiinfinite \textit{string-localized} vector potential
$A_{\mu}(x,e)$ or metric potential $g_{\mu\nu}(x,e)$ poses no problems i.e.
the missing possibilities in the spinorial formalism can be filled with
string-localized field generators. These covariant string-localized
"potentials" associated to pointlike "field strengths" possess mild short
distance property (scale dimension is one).

For Wigner's \textit{third kind of matter} the only known systematic
construction is one which determines a continuous ($\alpha$-dependent)
\ family of intertwiners $u^{\alpha}(p,e)$ using their modular localization
properties \cite{MSY1}\cite{MSY}. In this way one obtains a continuous set
which depend in addition to the momentum $p$ on a spacelike unit vector
$e,~e^{2}=-1.$ It intertwines the Wigner transformation, which involves the
representation $D_{\kappa}$ of the noncompact $E(2)$ little group with the
covariance transformation law in $p$ and $e$ and leads to a string field whose
intrinsic stringlike extension can be seen by the appearance of a nontrivial
commutator if one string enters the causal influence region of the other
\begin{align}
&  D_{\kappa}(R(\Lambda,p))u^{\alpha}(\Lambda^{-1}p,e)=u^{\alpha}(p,\Lambda
e)\\
&  \Psi(x,e)=\left(  \frac{1}{2\pi}\right)  ^{\frac{3}{2}}%
{\displaystyle\int_{\partial V_{+}}}
d\mu(p)(e^{ipx}u^{\alpha}(p,e)\circ a^{\ast}(p)+\nonumber\\
&  \qquad\qquad e^{-ipx}\overline{u^{\bar{\alpha}}(p,e)}\circ a(p))\nonumber\\
&  \left[  \Psi(x,e),\Psi(x^{\prime},e^{\prime})\right]  =0\text{,}%
~x+\mathbb{R}_{+}e~><~x^{\prime}+\mathbb{R}e^{\prime}\nonumber
\end{align}

That certain objects do not admit a pointlike presentation is not limited to
these third kind of Wigner (infinite spin) representations. The d=1+2
\ \textquotedblright plektons\textquotedblright\ (particle associated to braid
group statistics) are particles whose field theoretic description requires
spacelike strings \cite{Any}. By forming charge-neutral bilinear composites
one descends to compactly localizable observables. Last not least the
necessity from renormalizability of calculating with string-like potentials
instead of point-like field strengths even when the content of a theory can be
described in terms of the latter requires a new perturbative technology
\cite{M-S}. The iterative Epstein-Glaser step from the set of n-point time
ordered correlators to n+1 for rectilinear semiinfinite strings is certainly
more involved and without paying attention to the string-localization (only
paying attention to the end point) the construction of time ordered
correlation functions would lead to insoluble infrared problems.

Interactions of ordinary matter are taking place in a compact spacetime
region, at worst they are quasilocal. This places a question mark on whether
such matter can interact at all with standard matter and whether it can
trigger A-H counters. Leaving the final answer to future more detailed studies
there remains the question of whether a pair of such string-particles could be
created from a collision of ordinary matter. This should be possible if, as in
the case of 3-dimensional anyons \cite{Any}, the strings are somewhat
fictitious analogous to cuts in Riemann surfaces. Otherwise i.e. if the
asymptotic directions remain visible, causality should prevent the creation of
a pair of strings by colliding ordinary matter. In this case the mechanism for
darkness is kinematical and already occurs on a local level unlike the
previously mentioned hypothetical B-F subterfuge of the Aks theorem via the
breakdown of crossing. For the free field model at hand there are no
point-local fields which are quadratic in the creation/annihilation operators
as the following calculations shows.

It is important to know whether string-localized operators which applied to
the vacuum state create "one-string" states generate algebras which possess
point-like generated subalgebras.

The most general field which is quadratic in the annihilation/creation
operators and transforms as a scalar is of the form
\begin{align}
B(x)  &  =\int\int_{\partial V}d\nu(k)d\nu(l)d\mu(p)d\mu(q)\nonumber\\
&  e^{i\left(  p+q\right)  x}u_{2}(p,q)(k,l)a^{\ast}(p,k)a^{\ast
}(q,l)\label{B}\\
u_{2}(p,q)(k,l)  &  =\int d^{2}zd^{2}we^{i(kz+lw)}F(B_{p}\zeta(z)\cdot
B_{q}\zeta(w))\nonumber
\end{align}

The coefficient function must have the form in the third line; here
$\zeta(z)=(\frac{1}{2}(z^{2}+1),z_{1},z_{2},\frac{1}{2}(z^{2}-1))$ and
($z_{1},z_{2}$) is the Fourier transformed variables of the variable k
describing the space of the little group $E(2).$The intertwining function
$u_{2}$ clearly absorbes the complicated Wigner transformation of the
creation/annihilation operators and converts it into the simple transformation
of a scalar field.

In order to decide the nature of the localization property of $B(x)$ it is not
enough to consider its two-point function. According to the Kallen-Lehmann
representation its two-point function is automatically causal, but this only
means that the distribution-valued state vector $B(x)\Omega$ is
point-localized and implies nothing about the localization of the operator.
The string generated algebra would have compactly localized subalgebras in
case of existence of tensor fields which are relatively local to the string.
In case of our scalar bilinear field $B$ (\ref{B}) the answer to the
question:
\begin{equation}
\exists\ B\ s.t.~\left\langle q,l\right\vert \left[  B(x),\Psi(y,e)\right]
\left\vert 0\right\rangle =0,~x><y+\mathbb{R}_{+}e~? \label{loc}%
\end{equation}
is negative and this is best understood by comparing the contraction functions
with those for standard matter. By splitting off a plane wave exponential the
matrixelement in (\ref{loc}) only depends on the x-y difference. The Fourier
transform of this function is then polynomial in the Fourier momentum and this
leads to the spacelike vanishing. The presence of the$~z,w$ little-group
Fourier transforms in (\ref{B}) as well as in the definition of $\Psi(x,e)$
indicates a more complicated non-polynomial dependence which after Fourier
transform to the relative distance variable $x-y$ has no support properties at
all. A more pedestrian way to see this is to restrict the string and $B$ to
equal times. This situation cannot be improved by going from bilinear scalars
to tensors, or by generalizing from bilinear to 2n-linear expressions in the
$a^{\#}$ since by taking different matrixelements the contributions from
higher composites can be separated out and the same argument can be repeated.
With a probability bordering on certainty the algebras associated to Wigner
representations of the third kind do not contain pointlike localized
subalgebras and hence no local observables.

But such an illustration of complete absence of local observables in a free
field theory can hardly be taken serious as a realistic model for
darkness\footnote{The total lack of any relation to standard matter would only
leave the extremely speculative possibility that the origin of this matter is
inexorably linked to that of the still elusive quantum gravity.}. The question
is whether such objects can occur \textit{together with standard matter} as
part of the same theory id possible by more than the shared coupling to
gravity. A structural theorem by Buchholz and Fredenhagen shows that theories
with a mass gap may need semiinfinite stringlike generators for the algebras
but the application of these noncompact stringlike generators onto the vacuum
always creates states which can always be decomposed into pointlike localized
states; with other words even though the operator strings cannot be decomposed
into compactly localized parts inside the algebra, the states which they
create can be decomposed into irreducible massive Poincar\'{e}
representations. So such strings lead to superpositions of massive states of
the standard type which are not dark.

\section{Non-gravitational darkness}

In the introduction we mentioned a conjecture that massive matter which is
string-localized in the B-F sense is outside the range of the on-shell version
of Murphy's law (Aks theorem) may show very different behavior from standard
matter including darkness in the sense of vanishing production rates. The
existence of string-localized zero mass component of states, of which we
presented an interaction-free illustration, leads to a more startling kind of
darkness. The idea of relating indecomposable string-localized states with
"darkness" is based on the fact that the interaction with normal matter is at
least quasilocal i.e. takes place in compact regions. The quasilocality is a
compromise one has to make in order to describe counters which do not already
click in the vacuum. If one enlarges the local algebras to quasilocal by also
allowing operators which can be rapidly approximated by local ones, one finds
a dense subspace of quasilocal observables which have no vacuum polarization
and which are the objects in terms of which the A-H particle counters are constructed.

The lack of local subalgebras, in particular of quadratic pointlike local
composites, dampens hopes that such pure string-localized algebras could
account for DM. For whatever DM really consists of, it is certainly not inert
with respect to gravity and therefore it should contain a pointlike
energy-stress tensor. But the argument is not as cutting as it sounds if one
thinks in terms of quantum gravity. If one wants the metric tensor to be an
object of direct \textit{quantum} physical significance there is no way around
its string-localization since the \textit{quantum} potential of a (linearized)
point-like field strength $R_{\mu\nu\kappa\lambda}(x)$ is a string-localized
covariant symmetric tensor $g_{\mu\nu}(x,e).$

As in the analog s=1 situation of $F_{\mu\nu}(x)$ and $A_{\mu}(x,e)$ the
problem does not arise in the classical context because there is no quantum
imposition from Wigner's representation theory. Whereas in the s=1 case one
can argue that the vectorpotential is an auxiliary quantity which does not
have to be subjected to the quantum positivity requirements as long as one can
assure the positivity of a sufficiently large subsystem (which leads the
well-known gauge theory setting) this seems to be less palatable for the
$g_{\mu\nu}$ in quantum gravity. But then any classical modification of the
Einstein-Hilbert equation which involves other combinations of the metric
tensor would raise the issue of string-like localization if one starts from a
Minkowski background. So even before coupling DM the issue of localization of
gravity is not so clear.

Heeding advice from the gauge theory setting one expects the requirement that
interacting strings lead to point-like subalgebras to be very restrictive; in
fact any result different from an isomorphism between the point-like generated
subalgebra in the string-like formalism and the gauge invariant algebra in the
gauge setting would be deeply disturbing. Whereas the gauge setting trespasses
the positivity of QT but formally keeps the point-like covariant aspect of
vectorpotentials and finally arrives at local gauge-invariant fields by a
(BRST) cohomological descend, the approach based on string-localized
vectorpotentials leads to the same observable point-like generated subalgebra
but produces in addition very nonlocal operators which are the interacting
versions of massive matter fields which interact with the vectorpotential.

Taking again a lesson from QED one would like to think of the non-local
objects as corresponding in the gauge setting to nonlocal gauge invariant
operators as the delocalized electric charge-carrying operators. The
construction of these operators in the gauge setting is notoriously difficult
even in abelian gauge theories since it is not part of the formalism, but
rather requires a construction "by hand" \cite{Steinm}\footnote{For nonabelian
gauge theories the inability for doing such "by hand" constructions has been
an impediment to explore the physical content beyond the local gauge invariant
observables (gluonium,...) which left gluons and quarks in a very opaque
status (not very different from pDM).}. It is well known that electrically
charged particles are not Wigner particles but rather \textit{infraparticles}
i.e. stable entities which are surrounded by a soft photon halo and which
cannot be separated into a hard Wigner part and a soft photon
cloud\footnote{The energy-momentum spectrum is precisely that of "unparticles"
i.e. a accumulation of weight starting with a power singularty at the mass of
the charged particle which is submerged into the continuum. But charged
particles are the most visible "candles" in particle physics, and the
conjectured darkness of unparticles is a conceptual illusion caused by
insufficient knowledge of QFT and its history.}. The sharpest possible
localization of such infraparticles is semiinfinite stringlike and the
"minimal" generator (in the sense of lowest short distance dimension) is
formally given by the Jordan-Mandelstam formula of a Dirac spinor localized at
the endpoint of the string multiplied by an exponential function of a line
integral in terms of the vectorpotential; the difficult perturbative
renormalization status has been clarified by Steinmann \cite{Steinm}. The
impossibility to read these spacetime formulas a composite of spinor matter
and photon stuff complies with the impossibility to interpret an infraparticle
in terms of a mass-shell delta function and a continuum. This form of the
spectrum results from the quantum Gauss law which enforces a strong coupling
of infrared photons to the charge-current density (the Buchholz theorem
\cite{Bu}).

With string-localized vectorpotentials these delocalization aspects are built
into the formalism but there is a prize to pay at another place: one has to
understand a new renormalization theory in which the causal relations have to
be generalized to strings. Even though it would be enough to know the
perturbation theory for vectormesons with a string-fluctuation controlled by
localized fixed test function on d=1+2 de Sitter space (the range of spacelike
unit vectors) $A_{\mu}(x,g)=\int A_{\mu}(x,e)g(e)de,$ the renormalization
theory for string-localized fields requires to consider generic directions.
These problems are presently under investigation \cite{M-S}.

The QED illustration also focusses attention to the fact that it
string-localization in itself does not account for darkness. \textit{Our
Leitmotiv is rather that perfect darkness cannot be achieved in point-like
QFTs and that there are good reasons for expecting that there are
string-localized models with point-like subgenerators in which collision
between point-like generated matter does not lead to string-like generated
matter.}

This limits the ($s\leq1$) search for pDM to string-like localized mutually
interacting vectorpotentials with nontrivial pointlike generated subalgebras
i.e. to the string-localized reformulation of nonabelian gauge theories. It is
clear that if one succeeds to formulate renormalization theory for strings (a
very nontrivial conceptually and technically challenging step) one will have
gained a natural method for the construction of the nonlocal physical
operators which correspond to the above Jordan-Mandelstam string objects. The
main problem of the standard gauge formulation is that there is no natural way
to do this since the gauge formalism is focussed on local observables. In the
case of nonabelian interacting vectorpotentials one does not expect that the
local subalgebra generates the full physical space, rather there should be
lots of delocalized stuff in the physical Hilbert space. More concretely, the
nonlinear relation of the string-localized vector potentials to point-like
observables suggests that the former generate a space which genuinely contains
the one generated from the vacuum by the latter. This then could possibly lead
to an interacting model in which string-localized matter carrying positive
energy could be somewhat out of sight in the sense of this paper. We will come
back to this point in a somewhat broader context \cite{sch}.

Beside the intellectual pleasure derived from the problem whose solution
requires a profound knowledge of QFT, the main advantage of the linking dark
matter to structural properties of indecomposable string-localized fields is
that it allows to make a very precise experimental prediction: there should be
no possibility of pair creation from standard matter. In particular it would
contradict the predictions coming from supersymmetic extensions of the
Standard Model.

A null effect on the other hand would leave no alternative to strings as a
result of the Aks theorem for point-like fields based on the validity of the
crossing property. In that case the discoverer of the DM Fritz Zwicky and
Eugene Wigner, the protagonist of symmetry and of the first intrinsic
classification of particles (including irreducible representations which led
to the notion of string-like localization) may be linked \ by more than having
been scientists who wrote their important contributions at the same time.

As a theoretical physicist interested in conceptual problems, I always admired
Wigner's strict insistence in exploring known principles \textit{before} doing
mind games. Whereas the traditional way of valuating observations essentially
did not change since the time of Zwicky, the same cannot be said about modern
particle theory where the number of researchers following the intrinsic logic
of theoretical principles a la Wigner has decreased in favor of those doing
mind games.

.
\end{document}